# "Be Better Than You Need to Be": A-Level Physics Students' Rants, Resilience and Peer Pedagogy on TikTok

Wonyong Park

Southampton Education School, University of Southampton
1 University Road, Southampton SO17 1BJ, UK

w.park@soton.ac.uk

ORCID: 0000-0002-8911-5968

Abstract

Amid the growth of social media, students increasingly turn to short-form platforms for study support and advice. This study examines how A-level Physics is portrayed by student creators on TikTok, a popular platform amongst students for peer-to-peer educational guidance, information sharing, and collective knowledge building. Using reflexive thematic analysis of 57 TikTok videos via #alevelphysics, we explore how A-level Physics is characterised and what pathways to success are constructed within these informal discourses. The findings show that the subject is depicted as both intellectually demanding and emotionally taxing, with the examination system perceived as a flawed gatekeeper. Creators narrate their experiences through a blend of hyperboles, frustration, and resilience, offering sophisticated peer-generated pedagogies alongside cautionary tales. Recommendations emphasise strategic resource curation, sustained practice, systematic memorisation, and a mindset of deliberate over-preparation. These portrayals of subject demands and agentic responses resonate with prior research on the cultural construction of physics as hard and exclusionary, while also illustrating how affinity spaces on social media enable students to reframe difficulty through shared feelings and tactics. We discuss the value of student-led online discourses about high-stakes examinations for educators in supporting prospective and current physics students.

## Introduction

The rise of social media is rapidly reshaping the landscape of educational guidance, particularly for young people navigating the critical junctures of their academic journeys. Social media platforms such as TikTok have emerged as powerful arenas for peer-to-peer knowledge exchange, supplanting or supplementing traditional sources of advice like teachers, institutional websites, and formal careers services. A recent survey suggested that 48% of 3–17s who watch videos on video-sharing platforms use them to learn or for schoolwork (Ofcom, 2025). Within these dynamic digital environments, students share lived experiences, study strategies, and subject-specific perspectives, creating a vibrant ecosystem of informal guidance. This content, often perceived as more authentic and relatable, can present narratives about academic life that diverge significantly from official institutional discourse, offering a potent, user-generated counter-narrative to prospective students (Conde-Caballero et al., 2023; Jenkins, 2009; Raaper et al., 2024).

Such online spaces for peer academic interaction can be conceptualised through the lens of "affinity spaces," a term coined by Gee (2005) to describe sites where individuals gather around a shared interest or practice. Social media platforms facilitate the rapid formation of such spaces, allowing students with common academic goals or challenges to connect, collaborate, and co-construct knowledge outside of formal educational structures such as schools and colleges. These peer-led networks operate according to their own internal logics and norms, fostering a sense of community and shared identity (Gee, 2005). While they can provide invaluable support and demystify complex academic pathways, they also possess the capacity to perpetuate and amplify specific discourses, including potentially narrow or daunting portrayals of academic difficulty, success, and failure (Gee, 2004; Kirkpatrick & Lawrie., 2024), especially for subjects often seen as hard such as physics.

Despite the significance of narratives and discourses about physics, less is known about how they are reproduced, contested, and reconfigured within the contemporary affinity spaces of social media. This study addresses this gap by investigating the characterisation of A-level physics on TikTok, one of the most influential platforms for students today (Jerasa & Ura, 2025). Drawing on qualitative analysis of videos in which students discuss their A-level Physics experiences, the paper explores the messages and narratives that prospective students are likely to encounter—and that may shape their expectations, study choices and sense of the subject. Insights from this analysis can inform more effective instructional and guidance practices, as well as curriculum and assessment reforms aimed at making physics learning more accessible and meaningful for a wider range of students.

## Literature Review

### Physics at A Level

In the English education system, A levels (Advanced Levels) are subject-based qualifications typically taken by students aged 16–18. They are the main route into university study in the UK, though they are also valued by employers and vocational pathways. Students usually take three, and occasionally four, A-level subjects over two years. The content and assessment of each A level are set and administered by examination boards such as AQA, Pearson Edexcel, OCR, and WJEC Eduqas, which operate under regulation from the Office of Qualifications and Examinations Regulation (Ofqual), the government body that regulates qualifications and examinations in England. Assessment is mainly through final examinations at the end of the two years, though some subjects also include coursework or practical components, such as science practical endorsements. Grades are awarded on a scale from A* (highest) to E (minimum pass). Typically assessed through two to three written exam papers taken in May and June at the end of the two-year course, A levels are a key milestone that shapes students' post-school opportunities by influencing university admissions, career choices, and future study paths.

A-level physics, situated at the nexus of secondary and tertiary education in the UK, represents a particularly compelling case for examining peer interactions and discourses in social media. As a foundational subject for a wide array of STEM degrees and careers, its strategic importance is well documented (Royal Academy of Engineering, 2016; UCAS, 2025). Yet, it remains a subject with persistent challenges: in 2024, there were 43,114 A-level physics entries (10,026 female; 33,088 male), compared with 74,367 in biology and 62,583 in chemistry, and 107,427 in mathematics—illustrating relatively lower uptake and a marked gender imbalance (with about 23% female entries) and making it one of the most demographically skewed subjects in the English education system (Institute of Physics, 2021, 2023; Joint Council for Qualifications, 2024).

Research has illustrated the cultural construction of physics as an exceptionally challenging and gendered domain. DeWitt, Archer and Moote's large-scale work with over 13,000 Year-11 students shows how pervasive perceptions of physics as "hard," "mathematical," and implicitly "masculine" act as deterrents, especially for girls, shaping identities and a sense of "who physics is for" (DeWitt, Archer & Moote, 2019). Archer and colleagues further trace how students come to experience physics as "not for me," highlighting the role of identity, belonging and "science capital" in participation (Archer et al., 2020; Moote et al., 2020). In addition, evidence suggests that access to separate science GCSEs (triple science) significantly predicts A-level physics participation, with students taking separate sciences approximately three times more likely to continue to physics A-level (Plaister & Thomson, 2023). However, students from disadvantaged backgrounds remain grossly under-represented in triple science (Francis et al., 2025; The Royal Society, 2008). In a study by the Institute for Fiscal Studies (Cassidy et al., 2018), among girls predicted to achieve grades 7-9 in GCSE physics, 50% agreed that "I often worry that it will be difficult for me in physics classes," compared to 25% for mathematics. Similarly, 52% worried about getting poor grades in physics versus 38% in mathematics.

Research suggests that peer influence plays a major role in shaping adolescents' attitudes towards science (Breakwell & Beardsell, 2016). Oon, Subramaniam, Wong, and Abrahams (2024) found that secondary students in Singapore and Hong Kong perceived their close school friends as viewing physics as "uncool," whereas students in England expressed the opposite view. Such peer influence is significant because it shapes students' perceptions of their own abilities and can subtly influence their decisions about pursuing physics studies further (Breakwell & Beardsell, 2016; Simpson & Oliver, 1990; Vedder-Weiss & Fortus, 2013). Such a close relationship between peers and views of physics suggests the potential importance of a broader, virtual peer group that students are increasingly engaging with in online spaces. This is particularly relevant given that digital natives spend multiple hours daily online across platforms like Instagram, TikTok, and YouTube, where they engage with both academic and social content.

Assessment trends and wider public discourse further shape how physics is perceived and experienced. In the 2023–2025 period, concerns over exam difficulty and quality control gained sustained attention. In June 2024, for example, a widely shared online debate centred on a paper described by many candidates as "unreasonably difficult," while in June 2025, exam board OCR issued a formal apology for multiple errors in A-level Physics papers (Hodgson, 2025; Norden, 2024). Such episodes feed into ongoing perceptions of the subject as both challenging and, at times, unfairly assessed—sentiments that can influence both current learners' motivation and prospective students' decisions. Such contexts underscore the timeliness of examining not only participation patterns but also the ways in which students themselves make sense of physics in their everyday contexts. In an era where students' affinity spaces increasingly extend into social media environments such as TikTok, understanding how learners collectively frame subject difficulty, fairness, and strategies for success offers valuable insight into the lived realities of a high-stakes subject such as A-level Physics.

**Student Experiences and Agency in High-Stakes Exam Contexts**

High-stakes examinations such as A-levels represent pivotal moments in students' educational journeys, where individual performance can determine academic progression, career opportunities, and future life trajectories. Across diverse settings, and particularly in STEM, such assessment environments are consistently associated with heightened stress and anxiety (e.g., Putwain, 2023; Rozek et al., 2019). Effective coping and "academic buoyancy" can buffer some of the adverse effects of test anxiety on engagement and performance (Putwain, 2023; Putwain et al., 2015). Many students also routinely seek help and camaraderie from peers for both instrumental and emotional support. At a system level, while examination boards and subject associations do respond to candidate concerns, as seen in widespread reactions to an "unreasonably difficult" A-level Physics paper in 2024 and OCR's apology for errors in A-level Physics papers in 2025. (Hodgson, 2025; Norden, 2024), student voice remains under-represented in assessment reforms and policies, limiting opportunities to incorporate learners' situated expertise into system-level reforms (Barrance, 2019; Elwood, 2012).

Within this systemic constraints, students exercise agency by mobilising social, cognitive and emotional resources—often collectively. Research on socially shared regulation of learning highlights how groups coordinate goals, strategies and monitoring, distributing cognitive and emotional labour during demanding tasks—an analytic lens that aligns well with peer-led exam preparation practices (Järvelä et al., 2016). Students can actively exercise agency in high-stakes contexts by mobilising peer networks to share information, resources and study tactics. Teenagers use social networking sites to conduct peer-to-peer knowledge sharing that supports schoolwork and exam preparation (Asterhan & Bouton, 2017), while studies of secondary and post-secondary learners alike document how forums and group chats (e.g., WhatsApp) become informal "learning communities" where students pool expertise, co-regulate emotions and normalise difficulty (Dommett, 2019; Durgunoğlu & Yıldız, 2022; Mulyono et al., 2021; Pinchbeck, 2022).

From a sociocultural perspective, these spaces function as "affinity spaces" (Gee, 2004), enabling learners to craft shared identities as capable exam-takers and to translate tacit know-how (mark-scheme heuristics, time-management routines, revision schedules) into collective strategies. Such peer-led practices are linked to improved engagement and persistence, and they can mitigate the potential negative effects of high-stakes assessment by distributing cognitive and emotional labour across the group. These distributed practices are not merely incidental: meta-analysis suggests that structured support for regulation can enhance academic performance (Theobald, 2021), while the academic buoyancy literature shows that everyday resilience to routine academic setbacks offers protection against test anxiety during high-stakes periods (Putwain et al., 2023). In this light, peer networks (online forums, group chats, student-run study communities) constitute an important site of agency: they enable students to curate resources, align strategies with assessment demands, and co-regulate emotions under pressure—thereby mediating the high-stakes character of A-level assessment without waiting for system-level change.

While much of the existing work focuses on classroom-based interactions, online forums and messaging groups, little is known about how students publicly narrate and circulate exam-related advice within short-form video platforms where guidance, humour and identity work are interwoven. To address this gap, our study examines how A-level Physics is portrayed by student creators on TikTok and how agency is enacted through the sharing of strategies, routines and resources. Specifically, we ask: (1) How do student creators represent A-level Physics and its demands? (2) What pathways to success—study tactics, routines, resources—are articulated?

## Methods

### Data Corpus and Screening

This study employed a qualitative design, analysing publicly available TikTok videos. The search was conducted in August 2025 using the hashtag #alevelphysics, which yielded 299 videos. Inclusion criteria were: (a) explicit reference to studying or having recently studied A-level Physics; and (b) a first-person or peer-to-peer orientation. Videos were excluded if they: (a) were created by businesses or tutors (as determined from the video content or the creator's other videos); (b) primarily promoted products or services; (c) consisted mainly of music, images, or dancing with no or minimal speech; (d) focused chiefly on specific A-level Physics content or exam questions; or (e) were generally about physics learning without reference to A-level specifically.

Data were collected in August 2025, shortly after widely reported issues with A-level Physics papers in 2024–2025, which may have heightened the salience of assessment-related themes in TikTok posts. All videos with the hashtag #alevelphysics were searched without applying any time range restrictions. The 57 videos identified were posted between November 2022 and August 2025, spanning four calendar years: 1 in 2022, 8 in 2023, 14 in 2024, and 34 in 2025. While this pattern might suggest growing interest in the subject, it may also reflect the loss of older posts due to account closures, changes in privacy settings, or content removals.

Following screening, the final dataset comprised 57 videos centred on A-level Physics experiences. These covered a range of perspectives—advice, rants, reflections, and success stories—from current or recent students. Contrary to common perceptions of short-form platforms such as TikTok, a substantial number of videos contained rich accounts of students' lived experiences and peer advice. Most videos were between one and four minutes long, with creators often speaking fast from prepared scripts rather than improvisation. Videos were manually transcribed and entered into a spreadsheet for qualitative analysis. The resulting corpus totalled over 16,000 words. Although each video attracted user comments, these were typically too brief to be suitable for reflexive thematic analysis and were therefore excluded. A list of the analysed videos is provided in Appendix 1.

**Data Analysis**

To explore the dominant discourses surrounding A-level Physics, we employed reflexive thematic analysis—a qualitative method for identifying and interpreting patterns, or themes, within data that reflect people's lived experiences and perceptions in their social and cultural contexts (Braun & Clarke, 2006). Following Braun and Clarke's (2019) six-phase approach to reflexive thematic analysis, I began by immersing myself in the material, watching the videos and reading the transcripts repeatedly and memoing emergent impressions, as well as tonal distinctions between frustration, humour and encouragement, to situate early impressions in the platform's communicative style. Building on this familiarisation, I generated initial codes, capturing both semantic (explicit) meaning and latent (underlying) ideas

From this coded dataset I moved iteratively to theme development, clustering related codes into candidate themes that captured both the representations of A-level Physics (e.g., *emotional demands*, *intellectual challenge*) and pathways to success (e.g., *resource use*, *over-preparation*). During this phase, themes were refined, merged, or split. For example, initial separate ideas about "anxiety," "frustration," and "resilience" were consolidated into the more comprehensive theme "An Emotional and Psychological Ordeal". As a pragmatic check against idiosyncratic interpretation and overgeneralisation, I looked for resonance across at least three distinct videos while retaining notable deviant cases that complicated the dominant story. I then reviewed and refined these candidates, merging overlapping clusters and reworking or discarding thinner themes that lacked sufficient scope. Once the thematic map was finalised, each theme was given a concise, descriptive name that captured its essence. A detailed analysis was written for each, articulating its scope, boundaries, and relationship to the other themes and the two research questions.

Rigour of analysis was supported by a reflexive analytic stance, with contemporaneous memoing used to document decisions from initial coding through to theme development and naming (Charmaz, 2006). Interpretations were situated within the affordances and performance conventions of TikTok—brevity, humour, hyperbole, and stylised "rants"—so that meanings were read with sensitivity to the platform's communicative norms. The analysis focused on textual information, aligning with the requirements of reflexive thematic analysis, while recognising that delivery style and multimodal elements can shape interpretation. Potential counter-patterns were sought actively and preserved as deviant cases, enabling the refinement of thematic boundaries rather than forcing data into dominant narratives. For analytical transparency, the write-up provides illustrative excerpts alongside contextual details about the videos, allowing readers to assess the applicability of findings to other cohorts, subjects, or platforms.

## Results

**Representations of A-Level Physics**

The analysis of 57 video transcripts reveals a multifaceted and often highly emotional characterisation of the subject. While a minority frame it as manageable or even easy with the right methods, the overwhelming narrative portrays physics as a uniquely demanding and antagonistic force in a student's academic life. This representation is built upon three core themes: the subject as an intellectual antagonist defined by its extreme difficulty and conceptual leaps; the experience as an emotional and psychological ordeal that tests student resilience; and the examination system as a flawed and often hostile gatekeeper.

*An Intellectual Antagonist, with a Steep Learning Curve*

Across the dataset, A-level physics is frequently portrayed as "mythical", hostile, and often incomprehensible (Video 36). This characterisation is rooted in its perceived inherent difficulty, which many creators claim surpasses all other subjects. One student asserts, "this has to be the hardest a level of all time," citing anecdotal evidence that "people who did A-level further maths that said A-level physics is harder" (Video 20). This narrative of exceptionalism establishes physics as an outlier, a subject operating on a different plane of intellectual demand.

A key component of this antagonistic character is the "jump" from GCSE, which is described as uniquely severe. Unlike other subjects with more gradual learning curves, the transition to A-level physics is depicted as a sudden and disorienting shock. One creator warns prospective students, "the jump from GCSE physics to A-level Physics is 10 times harder. It. It is mad" (Video 45). This leap is not just in content volume but in conceptual abstraction. Another video elaborates on this, stating, "physics GCSE to AS [the first year of A level] is a big jumping content, and AS to A level is another quite big jumping content. I didn't find for maths or further maths or even economics, there was any sort of tangible jump" (Video 39; bracket added). This abrupt escalation contributes to the subject's intimidating reputation and is evidenced by high drop-out rates, as one student notes: "my physics class in year 12 we started with 30 students and by year 13 we were down to nine" (Video 42).

The content of A-level Physics itself is often framed as abstract, perplexing, and sometimes counterintuitive, further cementing its antagonistic role. Students voice frustration with conventions that seem arbitrary or needlessly complex. One creator, for example, rants about the units in astrophysics:

> What is astrophysics, bro? What is this? Like, I'm confused by a lot of the A-level content of physics, right? But this especially [...] why can't you just have half a degree? Or point of how many decimals of a degree? You don't need to turn it into a unit of time. It's a degree. Why is half a degree now 30 arc minutes or 1,800 arc seconds? [...] there is so much stuff that I'm perplexed about. Like, I'm actually stumped. (Video 16)

Here, the humour and incredulity sharpen the sense of physics as a subject that asks learners to navigate highly specific representational systems.

The link to mathematics is also identified as a core challenge, especially for those not taking A-level Maths. Although in principle "the hardest maths that can be included on a physics A level is GCSE maths" (Video 19), many note that being good at mathematics is an asset for A-level Physics. One student, faced with complex graphical analysis, exclaims, "We've got to like, learn both sides, draw graphs, get a gradient and like, learn it so it's a straight line graph. I don't even do Math A level, so how am I meant to do that" (Video 49). This sentiment is often expressed as a blunt warning to prospective students, with another creator stating that if you take physics, "Especially if you're not doing maths, you are finished" (Video 24).

*An Emotional and Psychological Ordeal*

The intellectual challenge of A-level physics is shown to inflict a significant emotional and psychological toll, framing the student experience as a ordeal of anxiety, frustration, and resilience. The transcripts are replete with visceral emotional language, describing students being "traumatised by magnetic fields" (Video 23), "humbled" (Video 22), or left feeling their soul has been sold for a good grade (Video 31).

A pervasive sub-theme is the deep-seated frustration that arises from a perceived disconnect between effort and understanding. The experience of being unable to comprehend material, even with the solution provided, is a common source of despair. One student powerfully articulates this:

Imagine I'm sat here doing a question, yeah, I go a bit stuck [...] so let me go and refer to the mark scheme. How are you gonna tell me? I look at the mark scheme as I still don't know what's going on. [...] The answer is laid out for me there. I still do not know what's going on. Do you not know how deep that is? (Video 36)

This sentiment highlights the intellectual helplessness that fuels the subject's reputation for being psychologically taxing. One creator recounts getting an E on a mock exam despite extensive revision, concluding, "That's how hard A-level Physics is" (Video 45), framing poor results not as personal failings but as an expected outcome of engaging with the subject. This leads to a normalisation of failure and struggle.

Consequently, resilience emerges as a critical, non-negotiable trait for survival. Students are advised that "you have to be resilient" because it is "very much a mindset thing," particularly when "you come out the exam and you think you've done horribly" (Video 6). Success is therefore not just about academic ability but about psychological fortitude. This includes managing pre-exam anxiety, with one student describing their fear The emotional strain can be so intense that some creators describe needing a "de-stress bash" as a "lifesaver" after the A-level exam — in one case, the multiple-choice section was said to have shattered the creator's dream "of having a loving wife in a nice house", delivered with semi-joking hyperbole (Video 52). The advice to "try and actually enjoy the challenge of it" (Video 6) is offered as a coping mechanism, an attempt to reframe the relentless struggle as a rewarding endeavour.

*The Examination System as a Flawed Gatekeeper*

The final theme characterises the examination system (including exam boards and mark schemes) as a flawed and often hostile gatekeeper. The difficulty of A-level physics is seen as being artificially inflated by unpredictable, overly specific, or outright erroneous exam papers. Students frequently direct their anger not just at the subject, but at the exam boards.

Exam papers are described in visceral terms as "garbage" (Video 24), a "real life nightmare" (Video 22), or the "worst paper I've ever sat in my whole life" (Video 46). A recurring complaint is the appearance of "rogue" questions that seem disconnected from the core syllabus or past paper trends. One student laments the randomness of a nuclear physics question: "Like working out Lambda and had number of throws of dice like with a probability like. We're not in stats, this is not maths" (Video 46). This unpredictability fosters a sense of unfairness, as if the examiners are deliberately trying to catch students out rather than test their knowledge. This perception is articulated in a direct, aggressive address to the exam board AQA:

> AQA, quick question about Physics Paper 2. Who the hell did you think was taking that test when you wrote that paper? Genuine, like genuine question. Do you think Einstein was taking that test? Do you think Isaac Newton was taking that test? [...] what possessed you to to write such a test, bro? (Video 47)

This raw, emotional quote captures the feeling that the exams are not designed for actual students but for historical geniuses, positioning the exam board as an unreasonable and out-of-touch adversary.

Beyond question design, logistical failures by exam boards exacerbate this feeling of hostility. One creator recounts an exam where the board "wrote the question wrong," invigilators forgot the equation sheet, and no extra time was given, leading them to call it "the worst paper I've ever done" (Video 25). Such errors undermine the legitimacy of the assessment process and add unnecessary stress. Furthermore, mark schemes are criticised for being "so specific and not intuitive like they might be in chemistry or maths" (Video 32), making it difficult for students to secure marks even when they understand the concepts. The constantly lowering grade boundaries are cited as definitive proof of the system's extreme difficulty, with one video tracking the A* [the highest grade] boundary's fall from 88% to 72% as evidence that the papers are becoming progressively harder (Video 1).

**Pathways to Success in A-Level Physics**

Despite the overwhelmingly negative characterisation of A-level physics as an intellectual and emotional hurdle, the TikTok student creators simultaneously construct a clear and consistent counter-narrative detailing pathways to success. This represents a powerful form of peer pedagogy, agentically constructed by students in direct response to their perception of the formal A-level physics system. Faced with a subject they often describe as hostile, opaque, and emotionally taxing, these creators take control by crowdsourcing and disseminating a shared, practical blueprint for survival and achievement. This student-led narrative is pragmatic and action-oriented, presenting high grades as achievable through disciplined and strategic effort rather than innate talent alone. Moving far beyond the generic "study hard" advice, this collective wisdom often provides battle-tested and specific methods. Four major themes define these pathways: building foundational understanding through strategic resource curation; the absolute primacy of relentless practice and application; developing systematic habits for memorisation and active recall; and cultivating a mindset of consistency and deliberate over-preparation.

*Building Foundational Understanding Through Strategic Resource Curation*

The first step on the pathway to success is the careful selection and use of online resources to build a solid conceptual foundation (see Table 1 for descriptions of the resources mentioned). Creators advise viewers not to rely on a single source but curate a bespoke library of tools, with specific platforms recommended for distinct purposes. YouTube emerges as the primary tool for initial learning and conceptual clarification. Channels like Science Shorts (mentioned in 7 videos), ZPhysics (mentioned in 6 videos), and Physics Online (mentioned in 3 videos) are repeatedly endorsed for their ability to "break down all of the A-level topics" in a way that is "so simple and satisfying" (Video 54). A popular strategy involves a two-step viewing process: "watch the whole playlist without writing anything. You're just gonna focus and listen [...] Then the second time [...] you're gonna write notes" (Video 35). This method separates passive absorption from active learning, ensuring a deeper understanding before moving to practice.

**Table 1.** A-level Physics support resources mentioned more than twice in TikTok videos

| Resource | Description | Videos Mentioned |
|---|---|---|
| Physics & Maths Tutor | A non-profit platform offering free revision notes, past papers, and worksheets for GCSE and A-Level students. | 5, 12, 13, 18, 37, 53, 54 (7 total) |
| Science Shorts | A YouTube channel providing concise video revisions of GCSE science topics, including full paper overviews in under an hour. | 4, 14, 22, 33, 35, 40, 48 (7 total) |

| ZPhysics | Created by a Head of Physics, this resource offers over 500 videos and online lessons for GCSE and A-Level physics. | 4, 10, 14, 32, 33, 54 (6 total) |
|---|---|---|
| Isaac Physics | A University of Cambridge project offering physics problems and support for students from GCSE level to university. | 3, 4, 8, 42 (4 total) |
| Physics Online | Create by a former Head of Physics, this resource features a large collection of free videos, premium plans, and live streams for GCSE and A-Level physics revision. | 4, 14, 33 (3 total) |
| UmuTech | Provides AQA A-Level Physics revision materials, including past papers, content packs, and mark schemes to practice exam questions. | 35, 43 (2 total) |

For developing problem-solving skills, Isaac Physics (mentioned in 4 videos) is consistently highlighted as an indispensable, albeit challenging, resource, mentioned in four videos. Its value lies in providing questions that are more demanding than typical textbook examples, thereby improving students' ability to "apply math in the context of sort of physics based questions" (Video 3). Creators acknowledge its difficulty, noting, "I know it is infuriating that they won't give you the answer," but this is framed as a positive, as "it makes it quite rewarding when you do get the answer" (Video 3). Isaac Physics is also recommended for its free mentoring schemes, which provide "weekly problem sets with maths and physics questions, which will really put help to push your math skills and physics problem solving skills" (Video 42).

Finally, websites such as Physics & Maths Tutor (PMT) (mentioned in 7 videos) and UmuTech (mentioned in 2 videos) are recommended for their vast repositories of practice questions organised by topic. These are positioned as essential for targeted revision once a foundational understanding is in place. One creator praises UmuTech as having "so many individual topic questions" that allow a student to "really hone into just thermophysics and perfect it" (Video 43). However, there's a cautionary note about over-relying on PMT topic questions, as doing so might exhaust the past paper questions needed for mock exams later (Video 53). The successful student, therefore, is one who navigates these resources strategically, using YouTube for concepts, Isaac Physics for problem-solving, and question banks for targeted drills.

*The Primacy of Practice and Application*

A central tenet across the entire dataset is that success in A-level physics is impossible without a relentless focus on practice and application. Mere content knowledge is deemed insufficient. As one creator starkly puts it, "You can know all the content and still get a D. Thing about physics is that you need to know how to apply this content, and the one way to do that is simply just practicing" (Video 15). The volume of practice recommended is immense. Creators advocate doing "loads and loads of past [exam] papers" (Video 6) and "as many multiple choice questions as you possibly can" (Video 3). The mantra is "questions, question, questions, questions. you cannot go wrong with doing questions" (Video 18). This includes utilising papers from the old specification, as "literally nothing has changed between those two" for many topics (Video 3). This high-volume approach serves to build fluency, reveal patterns in questions, and expose students to every possible variation of a problem.

Mastering practical skills is another critical sub-theme. This goes beyond knowing the experiments performed in class. Successful students must "know the other variations of that practical," as exams may feature an unfamiliar setup. The recommended method is to watch videos of all variants, because "it's so much easier if you're already familiar with them" (Video 3). Similarly, specific skills for the practical paper, like graph drawing, are broken down into meticulous detail: "You need to make sure your significant figures are correct [...] You need to make sure your graph takes up more than half the space [...] You need to make sure your line of best fit is steep" (Video 22). This focus on the minutiae of exam technique demonstrates that success is built on a foundation of disciplined, repeated application of knowledge to exam-style tasks.

### *Developing Habits of Memorisation and Active Recall*

The data also reveals that strategic memorisation is emphasised as a crucial and often overlooked pathway to securing marks. Creators push back against the idea that physics is "purely based on understanding", arguing that "there are quite a lot of things that you can memorise, and flashcards are genuinely essential for physics" (Video 3). These are particularly useful for "definitions and the diagrams", which are described as "such easy marks" that many students needlessly drop (Video 3).

Beyond definitions, memorising formulae that are not provided in the formula booklet is presented as a key exam hack. One creator outlines a highly specific routine for this:

> I'd memorise the ones I needed for a particular exam [...] but literally morning of the exam, I would go through them and make sure they're in my head, and then I would write them out on the top of the exam paper as soon as I could. And then they're there just in case I need to use them. (Video 3)

This tactical approach ensures that crucial information is readily available during the high-pressure exam environment, freeing up cognitive load for complex problem-solving.

Active recall is the method used to embed this memorised information. This involves moving beyond passively reading notes to actively retrieving information from memory. One creator advises the use of whiteboards to support memorisation: "You're gonna use the whiteboard for active recall. You're gonna write down everything you remember onto the whiteboard, and anything you miss out, you're gonna add after. And you're gonna keep doing this process until you've added everything onto your whiteboard" (Video 35). Another creator suggests generating flashcards and quizzes automatically from YouTube videos using an app, allowing them to "flick through them everyday just to learn all the content" (Video 29). These systematic habits ensure that both conceptual and factual knowledge is not just learned, but retained and readily accessible.

*Cultivating a Mindset of Consistency and Over-Preparation*

The final pathway to success involves cultivating a specific mindset grounded in consistency, discipline, and aiming beyond the minimum requirements. The demanding nature of the subject requires sustained effort over a long period. One student attributes their success to having "consistently revised, I think from about a month before the end of Year 12, we'll wait till the end of like Year 13" (Video 17). This consistency is built through daily habits, such as doing an hour of revision in the morning and a couple of hours after school. This approach normalises the heavy workload by breaking it into manageable chunks, making it feel less overwhelming.

A crucial element of the successful mindset is described as the pursuit of deep understanding over surface-level learning. It's not enough to know *what* happens; one must understand *why*. As one student explains:

> You're not memorising the content and you're not sort of learning the content, you're learning the why. You're trying to really understand the underlying mechanisms of everything that you're doing. [...] Because if you don't understand the why and you don't understand the how, then you're going to find it really, really difficult to be applying the knowledge. (Video 39)

This intellectual curiosity and commitment to first principles is what separates high-achieving students from those who merely memorise.

Finally, this mindset culminates in a strategy of deliberate over-preparation. Given the unpredictability of exams and the effects of pressure, students are advised to aim for a grade significantly higher than their target. One creator, reflecting on narrowly missing a top grade due to the stress of the day, offers this powerful advice as the core philosophy:

> "The exam pressure makes a big difference. [...] I made stupid mistakes even on the stuff that I know I knew. But anyway, just basically, like, Be better than you need to be. Do more work than you think you need to. Like, don't just stop because you need an A. Like, do it so that you could get a star. And then if not, you still get an A. You know what I mean?" (Video 17).

This philosophy acts as an insurance policy against the difficulty and stress of the final exams, encapsulating the ultimate pathway to success: building a margin of safety through relentless over-preparation. Such an appraoch is translated into concrete, often extreme, strategies for practice that aim for total mastery of all available materials. It is not enough to do a few past papers; students are advised to exhaust every possible resource to the point of complete saturation. The sheer volume recommended is immense, as one creator instructs peers to "do all the old syllabus papers. I'm talking, you know, Paper 1, Paper 2, Paper 3,

Paper 4, Paper 5 and paper 6 from your relevant exam board… then what you need to do is do the exam papers that are relative to your exam board… and as an addition to that, what you could do is do questions from other exam board" (Video 18).

## Discussion

Efforts to improve physics engagement and uptake in England have long confronted persistent issues: relatively low participation compared with other sciences, pronounced gender imbalance, and entrenched perceptions of physics as "hard" and "not for me" (DeWitt, Archer & Moote, 2019; Archer et al., 2020; Moote et al., 2020). These cultural narratives shape not only who chooses physics but also how those who do participate experience it in real time. In this context, student-generated portrayals of A-level Physics on social media provide an important window into how difficulty, identity, and success strategies are being reframed within the everyday media environments of young people.

One unique feature and strength of TikTok videos as a data source is that students address their peers directly, a form of communication that often elicits candid, unfiltered accounts of their experiences. Unlike interviews, which are shaped by the presence of a researcher, these peer-to-peer narratives on social media capture the subtle nuances as they occur in students' own social cultural spaces. Given the significant and growing influence of online peer interactions in shaping students' perceptions of learning, it is noteworthy—and potentially concerning for physics educators—that several TikTok videos, even accounting for the platform's often hyperbolic style, actively discourage viewers from taking A-level Physics with strong language, receiving numerous reactions:

> "Yo, listen, any future Year 12 students yet thinking of taking A-level Physics? Don't you. Take it for me. Do not do it. Because this subject right here is a flipping mythical [...] If you're thinking of taking A-level physics, please don't do it. Take it from me." (Video 36) (46,000+ likes, 1,000+ comments)

> "If you are between the ages of 15 to 16 and you have chosen A-level Physics, I recommend you drop it right now. I don't care if you got a grade on your mock, I don't care if you got a grade ten. If you're not doing engineering or physics in uni, drop it. Believe me. Look at my eyes. I have little bags under my eyes from four hours of sleep and revising from the whole of Year 13, just for this paper to be garbage." (Video 24) (3,000+ likes, 100+ comments)

Our analysis shows how creators on TikTok narrate A-level Physics as both forbiddingly difficult and, with the right approach, practically manageable. These portrayals align with Jenkins's (2009) concept of participatory culture and Gee's (2004, 2005) notion of affinity spaces, both of which emphasise how audiences co-create and circulate knowledge within informal, interest-driven networks. In our corpus, rants about abstraction or unfair assessment often pivot into "here's what worked for me," exemplifying the move from individual expression to collective knowledge-sharing. The short-form, performative nature of TikTok does not merely transmit advice—it shapes it into a vernacular of humour, hyperbole, and trend-aligned sounds, embedding guidance within familiar youth media practices. This scaffolding makes the content feel peer-authentic and socially embedded, echoing evidence that video-sharing platforms have become routine spaces of learning for young people (Ofcom, 2025). These spaces enable students to articulate a repertoire of practices—relentless practice, strategic resource curation, systematic memorisation—that collectively frame persistence as a rational and identity-consistent choice.

The way difficulty is narrated resonates strongly with literature on the cultural construction of physics as "hard" and implicitly masculine, and on the centrality of identity to participation decisions (DeWitt, Archer & Moote, 2019; Archer et al., 2020; Moote et al., 2020). Our creators reproduce the "hardness" trope—via talk of counter-intuitive conventions, "no banker" questions, and step-changes from GCSE—but they also offer counter-narratives that relocate the problem from innate ability to technique, routine and resource fit. In that sense, the TikTok portrayals both mirror and complicate earlier findings: they keep difficulty firmly in view, yet attach to it a repertoire of practices that make persistence feel rational and identity-consistent.

Practically, the findings can help educators to rebalance classroom focus to more explicitly address the areas students identify as critical for survival and success. Given the perception of the examination system as a "flawed gatekeeper". This calls for reforming the system in a way that ensures students' trust in the system. In the meantime, teachers can devote more time to demystifying assessment. This can involve dissecting mark schemes, explaining the rationale behind seemingly "rogue" questions, and teaching the metacognitive skills needed to tackle unfamiliar problems. At the same time, teachers should frame the student-led "be better than you need to be" ethos within paced, sustainable routines—regular short sessions, planned breaks, and protected sleep—to mitigate burnout and test anxiety, which are exacerbated by cramming and poor rest.

For prospective students, the implications are crucial for recruitment and retention. When speaking to GCSE classes, teachers can present a more authentic and empowering picture of A-level Physics—one that honestly acknowledges the challenges students will see discussed online but immediately pairs them with the clear, actionable pathways to success that current A-level students advocate for. This approach arms prospective learners with a realistic understanding and a practical plan, transforming the narrative from "not for me" to one of informed, agentic choice.

## Limitations and Future Research

One limitation of this research stems from its focus on the textual content of the video transcripts. Consequently, several core aspects of digital communication on TikTok, such as audience engagement metrics (e.g., user comments) and non-verbal, performative content—like videos where creators sing or dance—were not included in the analysis except when required to understand the context. This boundary was a consequence of employing a reflexive thematic analysis method, which was mainly designed to interpret text. These multimodal and interactive elements represent an increasingly significant form of self-expression and community-building among young people, highlighting a vital area for future work. Research that utilises methods capable of capturing these complex layers, such as multimodal discourse analysis or approaches informed by performance studies, supported by creator or user interviewees, can shed light on further nuances and details that were not captured by textual analysis. Such studies would provide a more holistic and nuanced understanding of students' multifaceted emotional and social responses to the culture of high-stakes examinations like A-level Physics.

## Data availability statement

All the data analysed for this study were retrieved from TikTok and are available on the platform at the time of writing. The creators might take the videos down later.

## Ethical statement

All videos were sourced from public TikTok posts. Transcripts were anonymised at the point of analysis and write-up, with no creator names, handles, or other traceable identifiers retained. Ethical approval for the study was granted by the author's university.

**Appendix 1.** Summary of TikTok videos about A-level Physics

| Number | Short Description |
|---|---|
| 1 | A video tracking the declining grade boundaries for A-level physics, showing the A* requirement dropping from 88% in 2019 to 72% in 2024. |
| 2 | A guide on how to improve from an E to an A, broken down into understanding concepts via real-life scenarios, remembering with Post-it notes, and reviewing with practice questions. |
| 3 | An A* student shares their secrets, highlighting the overlooked importance of memorizing definitions and formulas, using Isaac Physics for math skills, and mastering all practical variations. |
| 4 | A detailed list of recommended resources including Isaac Physics, Physics Online, specific YouTubers like Science Shorts and Z Physics, and a WJEC-specific university revision guide. |
| 5 | A student claims physics is easy to revise, crediting their success to active participation in class, extensive past paper practice from PMT, and genuine interest. |
| 6 | A student advises doing lots of past papers, maintaining a resilient mindset against bad results, and trying to enjoy the subject's inherent challenge as a motivator. |
| 7 | A creator seeks feedback on a new OCR A-level Physics revision website they are developing, which is similar in style to the popular MadasMaths resource. |
| 8 | A creator outlines a step-by-step method to get an A*, including identifying weak topics with a spreadsheet, using Isaac Physics, and downloading free flashcards. |
| 9 | A detailed revision method that the creator developed is presented, involving attempting questions from ET Physics playlists before learning the content, then using the videos and textbooks to solve them. |
| 10 | A two-day cramming guide for physics, which involves watching ZPhysics videos, making structured notes, and then memorizing the mark scheme answers for identified weak topic areas. |
| 11 | A video arguing that A-level physics students are "jarring" because they constantly complain about how hard their subject is, claiming maths is the harder subject. |
| 12 | A creator offers last-minute advice, emphasizing using mock results to identify weak topics, not neglecting Paper 3, and doing extensive practice from past papers and PMT. |
| 13 | A student who improved from an E grade advises doing all old-spec past papers until they are completely familiar and seeking explanations to truly understand content. |
| 14 | An A* student shares tips, including using varied resources like YouTube, mastering calculations, memorizing formulas, and completing timed practice papers near exams. |

| 15 | A student who went from a C to an A* explains that physics is purely about application and advises starting exam question practice from day one. |
|---|---|
| 16 | A student rants about the confusing nature of astrophysics, specifically questioning why angles are converted into seemingly illogical units like arc minutes and seconds. |
| 17 | A student reflects on narrowly missing an A-star due to exam pressure, advising others to over-prepare and aim for a grade above their target. |
| 18 | A creator states the most important thing is doing questions, recommending all old-spec papers, current-spec papers, and even questions from other exam boards. |
| 19 | A video explaining the "problem" with physics degrees: A-levels don't require maths, so students are shocked by the high level of maths at university. |
| 20 | A student declares physics the hardest A-level of all time, arguing that the questions are incredibly complex and require more than just knowing the content. |
| 21 | A creator warns GCSE students not to take A-level physics, stating the jump is enormous and the content is like a different, alien subject. |
| 22 | A creator describes the practical paper as a "nightmare" but offers advice on mastering it, including understanding concepts and perfecting graphing skills for easy marks. |
| 23 | A student, while pre-learning nuclear physics, reflects on being "traumatised by magnetic fields" and notes that easy-looking topics often have deceptive questions. |
| 24 | A student passionately warns younger students to drop physics unless it's essential for their degree, citing the "garbage" exam papers and extreme stress. |
| 25 | A student recounts their frustration over an exam where the board allegedly wrote a question wrong and invigilators forgot the equation sheet, undermining the test's fairness. |
| 26 | A student explains their decision to drop A-level physics for biology to pursue a career in dentistry, prioritising required subjects and avoiding academic burnout. |
| 27 | A creator reassures scared biology and chemistry students that physics is just a much harder subject, before offering revision tips for their own exams. |
| 28 | A student expresses severe anxiety the night before Paper 2, feeling the content is harder, rushed, and that no topic feels easy or safe. |
| 29 | A promotional video for an app that claims to make physics easy by automatically generating notes, flashcards, and quizzes from a YouTube video link. |
| 30 | A student humorously complains about having to use practice questions from 1984, highlighting the lengths they go to for revision materials. |
| 31 | A creator argues physics is the hardest A-level, refuting claims it's "light" and stating that getting an A is like "selling your soul." |

| 32 | A student attributes physics' difficulty to its highly specific and unintuitive mark schemes, before recommending the YouTube channel ZPhysics for clear explanations. |
| --- | --- |
| 33 | A curated list of the best YouTube channels for A-level physics, including Alice does Physics, Physics Online, Science Shorts, Z Physics, and Primrose Kitten. |
| 34 | An introductory video for an app that turns YouTube summary videos into notes and flashcards, presented as a secret method for getting top grades. |
| 35 | A detailed revision method is presented, involving watching Science Shorts videos twice, using a whiteboard for active recall, and practicing questions on UmuTech. |
| 36 | A creator passionately warns prospective students against taking physics, describing its "mythical" difficulty and the frustration of not understanding even with the mark scheme. |
| 37 | A guide on how to get an A-star, advising students to rank topics, learn the worst ones first using YouTube, and practice extensively on PMT. |
| 38 | An introductory video for an app that creates quizzes from YouTube links, claiming it helped the creator get 96% on a physics test. Same creater as Video 34 but a different video. |
| 39 | A deep reflection on A-level physics, emphasising the significant jump from GCSE, the importance of understanding the "why," and the deceptive difficulty of MCQs. |
| 40 | A last-minute revision guide suggesting watching Science Shorts walkthroughs and studying mark schemes directly from PMT to understand how to answer questions. |
| 41 | A video explaining that the A-level system creates a false impression of physics by not requiring A-level maths, leaving university students unprepared for the mathematical rigor. |
| 42 | An advice video for new students highlighting the 70% drop-out rate in the creator's class and recommending the Isaac Physics mentoring scheme to develop skills. |
| 43 | A student recommends the website UmuTech for its large bank of topic-specific questions, ideal for honing in on and perfecting weaker areas before exams. |
| 44 | A student expresses their pre-exam anxiety, stating their biggest fear is not the paper's difficulty but panicking and underperforming under pressure. |
| 45 | A creator shares their story of getting an E grade on a mock exam despite extensive revision, highlighting the massive difficulty jump from GCSE. |
| 46 | An angry rant about a "rogue" exam question that involved dice probability in a nuclear physics context, which felt unfair and out of syllabus. |
| 47 | An aggressive and emotional rant directed at the AQA exam board after Paper 2, questioning who they thought was capable of answering such a difficult test. |
| 48 | A creator recommends watching walkthroughs on Science Shorts and persevering with difficult questions for as long as possible before looking at the mark scheme. |

| 49 | A student expresses extreme frustration after a difficult Paper 2 exam, describing it as one of the most challenging papers ever with no easy questions. |
| --- | --- |
| 50 | A post-exam reflection where a student describes the paper as "bizarre," feeling they failed to apply their knowledge correctly and ran out of time. |
| 51 | A compilation of student interviews after a difficult Paper 3, expressing confusion, despair, and humorous resignation about their performance. |
| 52 | A creator rants about the "bumbaclot" difficulty of the AQA Paper 2 MCQs, joking that their distress ball was a "lifesaver" during the exam. |
| 53 | A strategic warning against overusing Physics & Maths Tutor topic questions too early, as it exhausts the past paper questions needed for later mock exams. |
| 54 | A creator praises the YouTube channel Z Physics for its simple and satisfying breakdowns, recommending watching summary videos before starting a new topic. |
| 55 | A student complains that A-level physics is incredibly boring, particularly the "turning points" option, and that the experiments lack the excitement of other sciences. |
| 56 | A compilation of humorous, mock post-match style interviews with students after their physics exam, where they express defeat and confusion about the questions. |
| 57 | A post-exam reflection where the creator found the paper synoptic but was frustrated by the hard multiple-choice section which involved a lot of guessing. |